\documentclass[12pt]{article}
\usepackage{amssymb,amsmath,bm,mathrsfs}
\usepackage{graphicx}
\newcommand{\bmat}{\left(\begin{array}}
\newcommand{\emat}{\end{array}\right)}

\def\beq{\begin{equation}}
\def\eeq{\end{equation}}
\def\beqa{\begin{eqnarray}}
\def\eeqa{\end{eqnarray}}

\def\-{\hphantom{-}}

\def\s2{\frac{1}{\sqrt2}}

\def\beq{\begin{equation}}
\def\eeq{\end{equation}}
\def\beqa{\begin{eqnarray}}
\def\eeqa{\end{eqnarray}}
\def\ba{\begin{array}}
\def\ea{\end{array}}

\def\IF{\relax{\rm I\kern-.18em F}}
\def\II{\relax{\rm I\kern-.18em I}}
\def\IP{\relax{\rm I\kern-.18em P}}
\def\IC{\relax\hbox{\kern.25em$\inbar\kern-.3em{\rm C}}}
\def\IR{\relax{\rm I\kern-.18em R}}

\def\cp{{\cal P}}

\def\Dsl{\,\raise.15ex\hbox{/}\mkern-13.5mu D} %this one can be subscripted
\def\IZ{Z\kern-.4em  Z}

 \def\cp#1{\relax\ifmmode {\IP\kern-2pt{}_{#1}}\else $\IP\kern-2pt{}_{#1}$\=fi}
\begin{document}

%\catchline{}{}{}{}{} % Publisher's Area please ignore

%\markboth{authorname}{Paper Title}
\begin{titlepage}

\begin{center}
\begin{large}
{\bf Matrix Quantization of Turbulence}\footnote{Based on invited talks delivered at:
 Fifth Aegean Summer School, ``From Gravity to Thermal Gauge theories and the AdS/CFT Correspondance'', September 2009, Milos, Greece; the Intern. Conference
on Dynamics and Complexity, Thessaloniki, Greece, 12 July 2010; Workshop on ``AdS4/CFT3 and the Holographic States of Matter'', Galileo Galilei Institute, Firenze, Italy, 30 October 2010.  } 

\end{large}

\vskip1truecm

Emmanuel Floratos

\vskip1truecm

{\sl Department of Physics, Univ. of Athens, GR-15771 Athens, Greece\\
Institute of Nuclear Physics, N.C.S.R. Demokritos,
GR-15310, Athens, Greece\\
mflorato@phys.uoa.gr}

\vskip2truecm

\end{center}
%\maketitle

%\begin{history}
%\received{(to be inserted by publisher)}
%\end{history}

\begin{abstract}
Based on our recent work on Quantum Nambu 
Mechanics $\cite{af2}$, we provide an explicit
quantization of the Lorenz chaotic attractor through the 
introduction of Non-commutative
phase space coordinates as Hermitian $ N \times N $ matrices in $ R^{3}$. For the volume preserving
part, they satisfy the commutation relations induced by one of the two Nambu Hamiltonians,
the second one generating a unique time evolution.  Dissipation
is incorporated quantum mechanically
in a self-consistent way having the correct classical limit without
the introduction of external degrees of freedom.
Due to its volume phase space contraction it violates
the quantum commutation relations. We demonstrate that the Heisenberg-Nambu
evolution equations for the
Matrix Lorenz system develop fast decoherence to N independent Lorenz attractors.
On the other hand there is  a weak dissipation regime, where the quantum mechanical
properties of the volume preserving non-dissipative sector survive for long times.

\end{abstract}

%\keywords{Chaos, Turbulence, Strange Attractors, Nambu Dynamics, Matrix Quantization}

\end{titlepage}
%\begin{multicols}{2}
\section{Motivation}																							
\noindent The study of turbulence is a very difficult and interesting subject in various fields of science and engineering
   since it is very common in nature \cite{land}. In modern fundamental physics it has been discussed in extremes scales and  
   temperatures  ranging from cosmology for its important role in the determination of the distribution of 
   voids and the distribution
   of peculiar velocities of galactic clusters around them \cite{schild} to the thermalization process of 
   quark-gluon plasma in heavy ion collisions \cite{abm} as well as to the non-gaussianities 
   in the velocity spectrum of liquid helium vortex interactions \cite{sciver}. 
   
   We are interested at the question of turbulence in quantum dissipative systems. 
   There, dissipation deforms the non-commutative phase space to classical phase
    space and the basic 
   question is about the spectra of relaxation times to classical behavior. 
   From the point of view of dissipation the problem is equivalent to the one 
   of open Hamiltonian \cite{weiss} 
   or non-Hamiltonian  systems \cite{tar}. Moreover
   turbulent behavior presupposes a fluid structure for the system with a superimposed
   mechanism for the creation and annihilation of vortices \cite{land}. 
   
   The phase space noncommutativity 
   in Quantum mechanics at times $t=0$ is given by the Canonical Commutation Relations(CCR)    
  \beq
	\left[  Q_{i}, P_{j} \right] \ = \ \imath \hbar \delta_{ij}   \ \ \ \  \ \ \ t=0 \ \ \ \  
	\eeq
	where $ Q_{i}, P_{i} \ \ \ i=1,2\cdots $ are conjugate canonical pairs of position and momentum 
	 for a system with n degrees of freedom. 
	 In the Heisenberg picture the operators develop in time according to the
	  Heisenberg-Hamilton eqs. of motion
	\beqa
	\imath \hbar \dot{Q}_{i} \ &=& \  \left[ Q_{i} , H \right] \ \ \ \ \ \ i=1,\ldots , n \nonumber \\
	\imath \hbar \dot{P}_{i} \ &=& \ \left[ P_{i} , H \right] 
	\eeqa	 
  with the CCRs preserved in time. 
  
  If dissipation is added linearly by hand then the Heisenberg eqs. become \cite{weiss}
 
 \beqa
 \imath \hbar \dot{Q}_{i} \ &=& \ \left[ Q_{i} , H \right] \ \ \ \ \ \ \ \ \ \ \ \ \ \  \\
 \imath \hbar \dot{P}_{i} \ &=& \ \left[ P_{i} , H \right] \ \ - \ \ \imath\hbar \Gamma P_{i} \ \ \ \ 
 \ \ i=1,\ldots , n  \nonumber 
 \eeqa
 with the commutator decaying exponentially in time 
 \beq
 \left[ Q_{i}(t) , P_{j}(t) \right] \ = \  \exp (-\Gamma t ) 
 \ \ \  \left[ Q_{i}(0) , P_{j}(0) \right]
 \eeq
  
 One of the most well studied toy models,as mechanisms for the onset of chaos, is the Lorenz 
 chaotic attractor \cite{eck}. Through the work of Ruelle and Takens an explosion of 
 theoretical and experimental work followed on the general roads to turbulence through
 the study of strange attractors in dynamical systems \cite{viana}. 
 
 If one is interested in the problem of turbulence in quantum systems such as, liquid Helium, Bose-Einstein
 condensates(BEC) or the quark-gluon plasma, or more generally in the quantum regime of the Big Bang 
 and its later epochs, one must create a framework to quantize the Navier-Stokes equation as well as its relativistic
 generalization \cite{landau}. 
 
 In the present talk we address this question in the case of a toy model for turbulence , the Lorenz attractor.
 Nevertheless we can pose the problem for more general dynamical systems. We now proceed to  
 tackle the following issues \cite{tar}
  
  a) Is it possible to quantize volume preserving flows in fluid dynamics of the form
  \beq
  \frac{d x^{i}}{d t} \ = \ f^{i} (\vec{x}) \ \ \ \ \ \ \ i=1,\cdots,n
  \eeq
  for general non-Hamiltonian dynamical systems which upon introduction of dissipation lead to strange
  attractors?
  
  b) The same problem but with the inclusion of dissipation that renders the system chaotic with strange attractors.
  
  In the present talk, we will answer both questions in the affirmative by expressing incompressible flow
  equations in terms of the Nambu n-brackets with n being the dimensionality of the dynamical phase space
  $ x^{1}, x^{2}, \cdots , x^{n} $ . For the case of the Lorenz  attractor $n=3$.
  We will show how to quantize the volume preserving Nambu flow equations, which are associated to the Lorenz 
  attractors and as a second step 
  we consider also the inclusion of dissipation. For more information on the Lorenz and R$\ddot{o}$ssler  
  attractors as well as their equivalents we refer the reader to our recent work\cite{af3,axen}.
  
  The quantization scheme we propose is analogous to the standard Heisenberg quantization of classical
  mechanics leading to a unique time evolution for the system. As a concrete example we will treat the 
  quantization of the Lorenz
  strange attractor by N $\times $ N Hermitian matrices implementing our flow decomposition 
  methodology into nondissipative-dissipative
  components.   

%-----------------------------------------------------------------------% 
 \section{Nambu vs Hamilton's Mechanics}
%-----------------------------------------------------------------------%
 
 In Hamilton's mechanics phase space is an even dimensional 
 symplectic   manifold
 $ (M_{2n})$ \ \ $ n=1,2,\cdots $ thus posssessing the structure 
 of a  Poissonian manifold\cite{arn}. We consider two such examples:
 
 A) The Real Plane $M_{2} \ = \ R^{2}$ 
 with Poisson bracket
  \beq
   \{ f , g \} \ = \ \epsilon_{ij} \partial^{i} f \partial^{j} g \ \ 
   \ \  \ \ \ f,g \in C^{\infty} (R^{2})
  \eeq
  which for the coordinates $ x_{1}, x_{2} $ gives
  \beq 
  \{ x_{i}, x{j} \}  = \epsilon_{ij} \ \ \ \ \ i,j=1,2 
  \eeq
 
  Given a Hamiltonian $ H \in C^{\infty}(R^{2})$ 
  a flow vector field is defined 
  \beq
  L_{H} \ = \ \frac{\partial H}{\partial x_{2}}\frac{\partial}
  {\partial x_{1}} - \frac{\partial H}
  {\partial x_{1}} \frac{\partial}{\partial x_{2}} \ \ ,\ \  
  \dot{f} = L_{H} f   
  \eeq
  
  For different Hamiltonians, we have the integrability condition
  \beq
  \left[ L_{H_{1}} , L_{H_{2}} \right] \ = \ L_{\{H_{1}, H_{2} \} } \ 
  \ \ \ \ \ \ \ 
  \eeq
  
  B)   $ M= S^{2} $  the 2-dimensional sphere embedded isometrically in 
  $R^{3}$,  \ \ \ 
  
  $ x_{1}^{2} + x_{2}^{2} + x_{3}^{2} = 1 $
    \ \ \ \  $  i=1,2,3 $ 
  
  with Poisson bracket   
  \beq
  \{ f , g \}_{S^{2}} \ \ = \ \ \epsilon_{ijk} \ \partial^{i} f 
  \  \partial^{j} g  \ x_{k}  \ \ \ \ \
  \eeq
  
  For the coordinates $ x^{i}$ it gives the $ SO(3) $ algebra
  \beq
  \{ x_{i}, x_{j} \}_{S^{2}} \ \ = \ \ \epsilon_{ijk} x_{k} \ \ \ 
  \ \ \ i,j,k=1,2,3
  \eeq
  
  The corresponding Hamiltonian flows are, $ H \in C^{\infty}(S^{2}) $,  
  \beq
  L_{H} \ = \ \epsilon^{ijk} \ x^{i} \ \frac{\partial H}{\partial x^{j}} 
  \frac{\partial}{\partial x^{k}} \ \ \ \ \ \ \ \  i,j,k=1,2,3 
  \eeq
  with Lie algebra the area preserving diffeomorphisms of the sphere 
  $ \mbox{SDiff}(S^{2})$, 
  \beq
  \left[ L_{H_{1}} , L_{H_{2}} \right] \ = \ L_{\{H_{1} , H_{2} \}}
  \eeq
  
  With the choice of basis functions for $C^{\infty}(S^{2})$, 
  the spherical harmonics $ Y_{lm}(\theta,\phi) $,  
  we get the algebra of $ \mbox{SDiff}(S^{2})$ \cite{arn,hop}. 
  
  \beq
   \{ H_{lm} , H_{l'm'} \}_{S^{2}} \ = \ f^{m m'm''}_{ll'l''} \ 
   \ H_{l''m''} \ \ \ \ \ \ 
  \eeq
   
  Y.Nambu in 1973 \cite{nam} proposed a generalization of 
  Hamiltonian mechanics for volume preserving flows. 
  The simplest example
  is for the $ M=R^{3} $ case. 
   For a given set $ f,g,h \in C^{\infty}( R^{3}) $ 
   he introduced the 3-bracket
 \beqa 
   \{ f, g, h \} \ &=& \ \epsilon_{ijK} \ \partial^{i} f \ 
   \partial^{j} g \ 
   \partial^{k} h     \nonumber \\      &=&  \vec{\nabla} f 
   \cdot  ( \vec{\nabla}g \ \times \ 
   \vec{\nabla} h )
   \eeqa
   which satisfies the integrability equation-Fundamental Identity(FI)
 \beqa
   \{  f_{1} , f_{2} , \{ f_{3}  , f_{4} , f_{5}\} \} &=& \{ \{ 
   f_{1} ,  f_{2} , f_{3} \} ,
   f_{4} , f_{5} \} \nonumber \\ &+& \{ f_{3} , \{ f_{1} , f_{2} , 
   f_{4} \} , f_{5} \} +
   \{ f_{3} , f_{4} , \{ f_{1} , f_{2} , f_{5} \} \} ,
   \eeqa
   
   L.Takhtajan in \cite{tak} showed that the n-bracket defines 
   a hierarchy of 
   $n-1, \cdots, 2$ brackets.
   
   For the $n=3$ for example by fixing the third function h in 
   (15) we get 
 
   \beq
   \{ f , g \}_{h} \ = \ \{ f , g , h \} \ \ \ \ \ \ \ \ \ \ \ 
   \forall  f,g \in C^{\infty}(R^{3}) 
   \eeq
   which from (16) satisfies the relation
   \beq
   \{ \{ f_{1}, f_{2} \}_{h} \ + \ \{ \{ f_{2} , f_{3} \}_{h} , 
   f_{1} \}_{h} \ + 
   \ \{ \{ f_{3},f_{1} \}_{h} , f_{2} \}_{h} \ = \ 0  
   \eeq
   which is  the Poisson-Jacobi identity. In this way we may 
   obtain a Poisson bracket for any surface
   in $ R^{3} $,  which is defined by the level set function 
   $ h(x_{1}, x_{2}, x_{3} ) = constant $.
   
   If we restrict the variables $ x^{1}, x^{2}, x^{3} $ to be on 
   the surface h then rel. (17) defines on it a non-degenerate
   Poisson bracket. In effect this surface gets promoted to a 
   phase space with dynamics.
   As an example we consider the plane $ R^{2} $ embedded in $ R^{3} $.
   
   \begin{enumerate}
    \item The Plane  $ h = \alpha^{i} \dot x^{i}$ , \ \ \ \  
    $\alpha^{i} \in R $   
    
   \beq
   \{ x^{i} , x^{j} \}_{h} \ = \ \epsilon^{ijk} \alpha^{k} 
   \eeq
    
   \item The Sphere $   x_{1}^{2} + x_{2}^{2} + x_{3}^{2}=1 $  , \ 
   \ \ \ $ S^{2} \in R^{3} $
	 
	 \beq
	 \{ x^{i} , x^{j} \}_{S^{2}} \ = \ \epsilon^{ijk} x^{k} \ \ \ \ \ \ \  
	 SO(3) \  on \ S^{2}
	 \eeq
	 \end{enumerate}
	 
	 Nambu introduced his mechanics for volume preserving flows in 
	 $ R^{3}$ by specifying 
	 $H_{1}, H_{2} \in C^{\infty} (R^{3})$ two Hamiltonians inducing  
	 the following eq. of motion
	 \beq
	 \dot{x}^{i} \ = \  \{ x^{i}, H_{1}, H_{2} \}, \ \ \ \ \ \ \ \ \ 
	 \ \ \ i=1,2,3
	 \eeq
	 or equivalently
	 \beq
	 \dot{\vec{ x}} \ = \ \vec{\nabla} H_{1} \ \times \ \vec{\nabla} H_{2}
	 \eeq
	 
	 These equations imply the conservation of  the ``\,Hamiltonians\,''
	 \beqa
	 \dot{H}_{1} \ &=& \ \dot{\vec{x}} \ \cdot \ \vec{\nabla}H_{1} \ = 
	 \ 0 \ \ \ \ \ \ \nonumber \\
	 \dot{H}_{1} \ &=& \ \dot{\vec{x}} \ \cdot \ \vec{\nabla}H_{1} \ = \ 0
   \eeqa
   The ideal example right from Nambu's paper is the Euler's top
   
   \beqa
   H_{1} \ \ &=& \ \ \frac{1}{2} \left( l_{1}^{2} \ + \ l_{2}^{2} \ + 
   \ l_{3}^{2} \right) \ \ \ \ \ \nonumber \\
   H_{2} \ \ &=& \ \ \frac{1}{2} \left( \frac{l_{1}^{2}}{I_{1}} \ + \ 
   \frac{l_{1}^{2}}{I_{1}} \ + \ \frac{l_{1}^{2}}{I_{1}} \right) 
   \eeqa
   with equations of motion
   
   \beqa
   \dot{l}_{1} \ &=& \ \left( \frac{1}{I_{2}} \ - \ \frac{1}
   {I_{3}}  \right) l_{2} \ l_{3} \ \ \ \ \ \ \ \nonumber \\
   \dot{l}_{2} \ &=& \ \left( \frac{1}{I_{3}} \ - \ \frac{1}
   {I_{1}} \right) l_{3} \ l_{1} \ \ \ \ \ \ \ \nonumber \\
   \dot{l}_{3} \ &=& \ \left( \frac{1}{I_{1}} \ - \ \frac{1}
   {I_{2}} \right) l_{1} \ l_{2} 
   \eeqa
   The generalization of the Hamiltonian flow vector fields are 
   the flow vector fields 
   \beqa
   L_{H_{1},H_{2}} \ \ &=& \ \epsilon^{ijk} \ 
   \frac{\partial H_{1}}{\partial x_{i}} \ 
   \frac{\partial H_{2}}{\partial x_{j}} \ \frac{\partial}
   {\partial x_{k}} \ \ \ \ \ \ \ \ \ \ \nonumber \\
   &=& \ \ \left( \vec{\nabla} H_{1} \ \times \ \vec{\nabla} H_{2} 
   \right) \cdot \vec{\nabla} \ \ = \ \ \ \vec{v} \cdot \vec{\nabla}
   \eeqa
   where
   \beq
   \dot{\vec{x}} \ = \ \vec{v} \ = \ \vec{\nabla} H_{1} \  
   \times \   \vec{\nabla} H_{2} \ \ \ 
   \eeq
   which are volume preserving and therefore describe flows 
   of incompressible fluids. 
   The question, which we pose now, is whether the inverse 
   also holds. Namely if for any flow
   \beq
   \dot{\vec{x}} \ = \ \vec{v}(\vec{x})
   \eeq
   which is incompressible
   \beq
    \vec{\nabla} \cdot \vec{v} \ = \ 0
    \eeq
    there exist two ``\,Hamiltonians\,'' $ H_{1} , H_{2} $ with
    \beq
    v^{i} \ = \ \{ x^{i} , H_{1} , H_{2} \} \ \ \ \ \ \ \ \ \ \ \ \ \ 
    \ \ \\
    \eeq
    Since a long time ago \cite{lamb}  the answer to this problem 
    has been settled  to 
    the affirmative locally in $ R^{3}$. Indeed for any 
    incompressible vector flows  $ \vec{v} $ , 
    there is a vector potential $ \vec{A} $ such that  
    $ \vec{v} = \vec{\nabla} \times \vec{A}$  which, moreover,
    satisfies the Clebsch-Monge decomposition (CM- stream potentials),
    \beq
    \vec{A} \ \ = \ \ \vec{\nabla} \alpha \ + \ \beta \vec{\nabla} \gamma 
    \eeq
    from which we can deduce that
   \beq 	
   \vec{v} \ = \ \vec{\nabla} \beta \ \times \ \vec{\nabla}\gamma  \ \ 
   \ \ \ \ 
   \beta= H_{1} , \gamma=H_{2} 
   \eeq
    
    The corresponding infinite dimensional algebra of volume 
    preserving diffeomorphisms in the 
    Clebsch-Monge gauge is \cite{af1}
    \beq
    \left[ L_{H_{1}, H_{2}} \ , \ L_{H_{3}, H_{4}} \ \right] \ \ = \ 
    \ L_{ \{ H_{1}, H_{2}, H_{3} \}, H_{4} } \ + 
    \ L_{H_{3},\{ H_{1}, H_{2}, H_{4}\}} \ \ \ \ \ \ \ \ \ \ \ \ \ 
    \eeq
    So the full structure of the infinite dimensional group is 
    determined by the 3-bracket algebra of 
    a basis of functions $ H_{i} \in C^{\infty}(R^{3}) , \ \ \ 
    \ i=1,2,\cdots $ \cite{af2}
    \beq
    \{ H_{i} , H_{j} , H_{k} \} \ = \ f_{ijk}^{l} \ H_{l} \ \ \ \ \ \  
    \ \ \ \ \ \ \ \  i,j,k,l=1,2,\cdots
    \eeq
    
    As an example we present the Nambu 3-algebra on $ S^{3}$.
    
    Consider $S^{3}$ isometricallly embedded in $R^{4}$ 
    \beq    
     x_{1}^{2} + x_{2}^{2} + x_{3}^{2} + x_{4}^{2} = 1  
    \eeq 
     where $ x_{i} \in R ,\ \  i=1,2,3,4 $ 
    The Nambu bracket on $ R^{4}$                             
    \beq
    \{ x_{i} , x_{j} , x_{k} , x_{l} \}_{R^{4}} \ = \ \epsilon_{ijkl} 
    \ \ \ \ \ \ \ \ \ \ \ i,j,k,l=1,,2,3,4
    \eeq
    induces a 3-bracket on $S^{3}$. 
    Define : $ \forall \ f_{i} \ \in \ C^{\infty} \ (S^{3}) \ \ \ 
    \ \ i=1,2,3 $
    \beq
    \{ f_{1} , f_{2} , f_{3} \}_{S^{3}} \ = \ \epsilon^{ijkl} \ x^{i} 
    \ \partial^{j} f_{1} \ 
    \partial^{k} f_{2} \ \partial^{l} f_{3} 
    \eeq 
    which on the $R^{4}$ coordinates gives
    \beq
    \{ x_{i}, x_{j} , x_{k} \}_{S^{3}} \ = \ \epsilon_{ijkl} \ x_{l} \ 
    \ \ \ \ \ \  i,j,k,l=1,2,3,4
    \eeq
    This is the celebrated Baggert-Lambert algebra for
    $  SO(4)$ \cite{bl}.
    For any $ f,g \in C^{\infty}(S^{3}) $ we consider the flow 
    vector field
    \beq
    L_{f,g} \ \ = \ \ \{ f , g , \cdot \}_{S^{3}} \ \ \ \ \ \ \ \ \ \ \ 
    L_{f,g} h \ = \ \{ f , g , h \}_{S^{3}}
    \eeq
    and thus the Nambu equations  on $S^{3}$ are given as :
    \beq
    \dot{x}^{i} \ \ = \ \ L_{H_{1},H_{2}} \ x^{i} \ \ = \ \ 
    \{ x^{i} , H_{1} , H_{2} \}_{S^{3}} \ \ \ \ \ \ \ \ \ \ 
    \forall \ H_{1}, H_{2} \  \in \ C^{\infty}(S^{3})  
    \eeq
    \newpage
%------------------------------------------------------------------------
 \section{ The Lorenz Attractor as a Dissipative Nambu Mechanics}
%------------------------------------------------------------------------   
    We are not going to repeat the history of the famous 
    Lorenz-Saltzman system \cite{sal,lor} of nonlinear equations,
    which is the Galerkin trancation of the Fourier modes of 
    the temperature gradient and stream 
    potential for the Rayleigh-Benard experiment. We will
    nevertheless present their dynamical system of three variables
    \beqa
    \dot{X} \ &=& \ \sigma \ ( Y \ - \ X ) \ \ \ \ \\
    \dot{Y} \ &=& \ X \ ( r \ - \ Z ) \ - \ Y \ \ \ \ \ \nonumber \\
    \dot{Z} \ &=& \ X \ Y \ - \ b \ Z \ \ \ \ \ \nonumber
    \eeqa
    with
    $ \sigma $ being the  Prandl  Number, r  the  relative Reynolds number 
    and b is the aspect ratio. 
   
    The system has as critical points
    \begin{enumerate}
    \item $X_{1} \ = \ Y_{1} \ = \ Z_{1} \ = 0$
    \item $X_{2} \ = \ Y_{2} \ = \ \sqrt{ b ( r - 1) } \ , \ \ 
    Z_{2} = r - 1$ 
    \item $X_{3} ,\ Y_{3} \ = - \sqrt{b(r-1)} \ , \ \ Z_{3} \ = \ 
    r - 1 $  
    \end{enumerate}
    
    Using stability analysis for fixed $\sigma$ and b, their behaviour 
    is controled by the
    relative Reynolds number r. 
    Turbulent behavior starts for values of 
    \beq
     r = r_{H} \ = \ \sigma \  \frac{\sigma + b + 3}{\sigma - b - 1}    
    \eeq
    with
    $\sigma$ = 10 , \ \  $b = \frac{8}{3} $ ($\rightarrow 
    r_{H} = 24.73)$). Lorenz's choice is $r=28 > r_{H} $ in
    order to obtain chaotic  behaviour \cite{spa}
    where Hopf bifurcation appears with
    $ \vec{X}_{1}$ a saddle point with $\vec{X}_{2}$ and 
    $\vec{X}_{3}$  going repelling-unstable.
    
    In our recent work \cite{af3} we implemented a decomposition of 
    the Lorenz vector flow into a
    volume preserving (non-dissipative) and a dissipative components
    \beq
    \vec{v} \ = \ \vec{v}_{ND} \ + \ \vec{v}_{D} 
    \eeq
    The non-dissipative component gives a volume preserving flow 
    $ \vec{\nabla} \cdot \vec{v}_{ND} =0 $
    \beq
    \vec{v_{ND}} \ \ = \ \  ( \sigma Y , \  X (r - Z) , \  X Y )  
    \eeq
    with
    \beqa
    \dot{X} \ &=& \ \sigma Y \ = \ \{ X , H_{1} , H_{2} \}  \nonumber \\
    \dot{Y} \ &=& \ X ( r - Z ) \ = \ \{ Y , H_{1}, H_{2} \}  \ \ \ \\
    \dot{Z} \ &=& \ X Y \  = \ \{ Z , H_{1}, H_{2} \} \ \ \ \nonumber
    \eeqa
    with
    \beq
    \vec{V}_{ND} \ = \ \vec{\nabla} H_{1} \ \times \ \vec{\nabla} H_{2}
    \eeq
    
    We have identified  the conserved  Hamiltonians (see also \cite{nb})
    \beqa
    H_{1} \ &=& \ \frac{1}{2} \ \left( Y^{2} \ + \ ( Z - r )^{2} 
    \right) \ \ \ \ \ \ \nonumber \\
    H_{2} \ &=& \ \sigma Z \ - \ \frac{X^{2}}{2} \ \ \ \ \ \ \ 
    \eeqa
    
    The corresponding Poisson brackets on the surface $ \Sigma $, 
    $ H_{2}=$constant are
    \beqa
    \{ X , Y \}_{\Sigma} \ &=& \ \partial_{Z} H_{2} \ = \ \sigma  
     \ \ \ \  \nonumber \\
    \{ Y , Z \}_{\Sigma} \ &=& \ \partial_{X} H_{2} \ = \ - X \ \ \ 
    \ \nonumber \\
    \{ Z , X \}_{\Sigma} \ &=& \ 0 
    \eeqa
    
    By eliminating the variables Y,Z from eq.(45) we get     
    \beq
    \ddot{X} \ + \ ( H_{2} \ - \ \sigma r ) \ X \ + \ \frac{X^{3}}{2} \ 
    = \ 0 
    \eeq
    which is just the equation of motion for the one dimensional Anharmonic 
    Oscillator. Depending on the initial conditions
    we identify its two familiar symmetry phases as:
    \begin{itemize}
    \item $ H_{2} \ - \ \sigma r \ \geq 0 $  \ \ \ \ \ \ 
    Single well-Symmetric Phase 
    
    \item  $ H_{2} \ - \ \sigma r \ \leq 0 $  \ \ \ \ \ \ \ \ \ \ \ 
    \ Double Well-Broken Phase
    \end{itemize}
    
    By including dissipation we have
    \beq
    \dot{\vec{X}} \ = \ \vec{\nabla} H_{1} \ \times \ \vec{\nabla} 
    H_{2} \ - \ \vec{\nabla} D \ = \ 
    \vec{V}_{ND} \ + \ \vec{V}_{D} 
    \eeq
    
    where the dissipation potential D is given below
    
    \beqa
    H_{1} \ &=& \ \frac{1}{2}( (z-r)^{2} + y^{2} ) \ \ \ \ \nonumber \\
    H_{2} \ \ &=& \ \ \sigma Z \ - \ \frac{X^{2}}{2} \ \ \ \ \ \ \  \\
    D \ \ &=& \ \ \frac{1}{2} \ ( \sigma X^{2} \ + \ Y^{2} \ + 
    \ b Z^{2} )\ \ \ \nonumber 
    \eeqa 
    Now, of course, $H_{1}, H_{2} $ along with $ D $ are not any 
    more conserved. But still 
    $ H_{2} $ is a useful quantity since by eliminating Y,Z we find 
    the dynamical system  for 
    X, $ H_{2} $ 
   \beqa
    \ddot{X} \ + \ ( 1 + \sigma) \ \dot{X} + X ( \frac{X^{2}}{2} \ + 
    \ H_{2} \ - \ \sigma(r - 1) ) \ &=& \ 0 
    \ \ \ \nonumber \\  
    \dot{H}_{2} \ + \ b H_{2} \ - \ \sigma ( 1 \ - \ 
    \frac{b}{2\sigma} ) \ X^{2} \ \ = \ \ 0
    \eeqa
   
   The anharmonic oscillator moerover develops a damping term and since 
   $ H_{2} $ is not conserved, we obtain a sequence
   of symmetry breaking and symmetry restoring phases. By solving for 
   $ H_{2} $ we obtain the well 
   known Takeyama
   memory term in the anharmonic potential \cite{takeya1,takeya2} 
   \beq
   H_{2}(t) \ = \ H_{2}(0) e^{-b t} \ + \ \sigma (1 - 
   \frac{b}{2\sigma}) \ e^{-b t} 
   \int_{0}^{t} \ d\xi \ e^{b \xi} \ X^{2} (\xi) 
   \eeq
 %-------------------------------------------------------------------------%-----------------------------------------------------
\section{Matrix Quantization of the Lorenz Attractor}
%-------------------------------------------------------------------------%-----------------------------------------------------
   We will quantize the volume preserving part, in the Nambu form, of the Lorenz attractor system. 
   Firstly we write the Nambu equations, 
   \beqa
\dot{X} \ \ &=& \ \ \{ X , H_{1} \}_{H_{2}} \  =  \{ X , H_{1} , H_{2} \} \ \ \ \ \ \ \nonumber \\
\dot{Y} \ \ &=& \ \ \{ Y , H_{1} \}_{H_{2}} \ = \{ Y , H_{1}, H_{2} \} \ \ \ \ \ \ \   \\
\dot{Z} \ \ &=& \ \ \{ Z , H_{1} \}_{H_{2}} \ = \{ Z , H_{1} , H_{2} \}  \ \ \nonumber 
\eeqa

Using the Poisson algebra of equations (48).
\beqa
\{ X , Y \}_{H_{2}} \ \ &=& \ \ \sigma \ \ = \ \ \partial_{Z} \ H_{2} \ \ \ \ \ \ \ \nonumber  \\
\{ Y , Z \}_{H_{2}} \ \ &=& \ \ - X \ \ = \ \ \partial_{X} H_{2} \ \ \ \ \ \ \ \ \ \ \\
\{ Z , X \}_{H_{2}} \ \ &=& \ \ \partial_{Y} H_{2} \ \ = \ 0 \ \ \ \ \ \nonumber 
\eeqa

The quantization of the system is carried through the quantization of the Poisson algebra, 
i.e. by lifting it to commutators
with Weyl-ordered operators  
\beqa
\left[ \widehat{X} , \widehat{Y} \right] \ \ &=& \ \ \imath \hbar \sigma \ \ \ \ \ \ \ \ \ \ \ \nonumber  \\
\left[ \widehat{Y} , \widehat{Z} \right] \ \ &=& \ \ - \imath\hbar \widehat{X} \ \ \ \ \ \ \ \ \ \ \ \ \ \ \\
\left[ \widehat{Z} , \widehat{X} \right] \ \ &=& \ \ \ 0 \ \ \ \ \ \nonumber 
\eeqa

We must find either infinite dim. matrices or differential operators
that satisfy rel. (56)
The Quantum Nambu equations for the volume preserving part follow,

\beqa
\imath \hbar \dot{\widehat{X}} \ &=& \ \ \left[ \widehat{X} , \widehat{H}_{1} \right]_{H_{2}} \ \ \ \ \ \ \ \ \nonumber \\
\imath \hbar \dot{\widehat{Y}} \ &=& \ \ \left[ \widehat{Y} , \widehat{H}_{1} \right]_{H_{2}} \ \ \ \ \ \ \  \\
\imath \hbar \dot{\widehat{Z}} \ &=& \ \ \left[ \widehat{Z} , \widehat{H}_{1} \right]_{H_{2}}  \ \ \ \ \ \ \ \ \nonumber
\eeqa
where the $H_{2} $ index implies that we evaluate the commutators by using the algebra (56) which has as Casimir the 
second Hamiltonian $ \widehat{H}_{2} $( parabolic-cylinder phase space)
\beq
\widehat{H}_{2} \ = \ \frac{\widehat{X}^{2}}{2} \ - \ \sigma \widehat{Z}
\eeq
We obtain
\beqa
\dot{\widehat{X}} \ &=& \ \sigma \widehat{Y} \ \ \ \ \nonumber \\
\dot{\widehat{Y}} \ &=& \ - \frac{1}{2} \ \left( \widehat{X} \widehat{Z} + \widehat{Z} \widehat{X} \right) \ + 
\ r \widehat{X} \nonumber\\
\dot{\widehat{Z}} \ &=& \ \frac{1}{2} \ \left( \widehat{X} \widehat{Y} + \widehat{Y} \widehat{X} \right) 
\eeqa
from where we get the conservation of $ \widehat{H_{1}}, \widehat{H_{2}} $ and a 
unique time evolution
\beqa
\widehat{X}(t) \ &=& \ e^{-\frac{i}{\hbar} t H_{1}} \ \widehat{X}(0) \  e^{\frac{i}{\hbar} t H_{1}} \ \ \ \ \  \ \ \nonumber \\  
\widehat{Y}(t) \ &=& \ e^{-\frac{i}{\hbar} t H_{1}} \ \widehat{Y}(0) \  e^{\frac{i}{\hbar} t H_{1}}  \ \ \ \ \ \ \ \\
 \widehat{Z}(t) \ &=& \ e^{-\frac{i}{\hbar} t H_{1}} \ \widehat{Z}(0) \  e^{\frac{i}{\hbar} t H_{1}} \ \ \ \nonumber
\eeqa

By eliminating $ \widehat{Y}, \widehat{Z} $ from rels. (59) we get the anharmonic oscillator quantum eqs.
\beq
\ddot{\widehat{X}} \ + \ \left[ ( \widehat{H}_{2} \ - \ \sigma r ) \widehat{X} \ + \ \frac{\widehat{X}^{3}}{2} \right] \ = \ 0
\eeq

 The $H_{2}$ Casimir values determine superselection sectors for rel.(61) with $ H_{2} < \sigma r  (\mbox{Double Well})$, where
quantum tunneling appears. For $ H_{2} \geq \sigma r (\mbox{Single Well})$ 
anharmonic potential governs quantum dynamics. Both cases possess 
discrete spectra. 

For the non-dissipative part of the quantum Lorenz system, we could also work in the Schrondinger picture for the one
dimensional anharmonic potential.

%-------------------------------------------------------------------------%-----------------------------------------------------
\section{ Inclusion of Dissipation; Decoherence of the Quantum Lorenz Attractor}
%-------------------------------------------------------------------------%-----------------------------------------------------
The Quantum Lorenz equations including dissipation are \cite{af3} 

\beqa
\dot{\widehat{X}} \ &=& \ \sigma \left(\widehat{Y} - \widehat{X} \right) \ \ \ \ \ \nonumber \\
\dot{\widehat{Y}} \ &=& \ -\frac{1}{2} \ \left( \widehat{X}\widehat{Z} \ + \ \widehat{Z}\widehat{X} \right) \ + r \widehat{X} - \widehat{Y} \  \\
\dot{\widehat{Z}} \ &=& \ \frac{1}{2} \left( \widehat{Y} \widehat{X} + \widehat{X} \widehat{Y} \right) \ - \ b \widehat{Z} \nonumber
\eeqa

It is obvious that the quantum algebra(56) does not hold anymore for all times and $ \widehat{H}_{1}, \widehat{H}_{2}$ are not conserved. Still we can eliminate $ \widehat{Y}, \widehat{Z} $ and obtain a dynamical anharmonic system with quantum Takeyama memory term. The Schrondinger picture is not anymore convenient due to the time nonlocality of the memory term.
Qualitatively one expects that weak dissipation will induce a broadening of the discrete energy levels. Moreover for the case
of strong dissipation the quantum system is expected to collapse to the classical one. 

We can approximate the quantum system through the usage of finite dimensional $ N \times N $ Hermitian matrices for
the representation of $ \widehat{X}, \widehat{Y}, \widehat{Z} $. This is, for example, the case with a spin system with
$N=2s+1$. See, for example, the case in \cite{af1} of an Euler top
interpretation for the Lorenz attractor. Numerical experiments for the non-dissipative systems,  show integrable behaviour while for weak dissipation we obtain N weakly interacting Lorenz attractors. On the other hand for strong dissipation the quantum system quickly undergoes
decoherence to a classical system of N independent Lorenz attractors ( as $ \widehat{X}, \widehat{Y}, \widehat{Z} $ become 
asymptotically mutually commuting).

We evaluated numerically the time evolution of the N $\times $ N matrix 
Lorenz system with the standard parameter values
 $( r=28 , \sigma = 10, b= \frac{8}{3} )$. 
 
We found that the commutators of
$  \widehat{X}, \widehat{Y}, \widehat{Z} $ undergo rapid decoherence unless dissipation is weak.
 \beqa
\left[ \widehat{X}(t) , \widehat{Y}(t) \right]  \ &\stackrel{t \rightarrow\infty } \longrightarrow & \ \ e^{- \Gamma_{12} \cdot t}  \ \left[ \widehat{X}(0) , \widehat{Y}(0) \right] 
\nonumber \\  
\left[ \widehat{Y}(t) , \widehat{Z}(t) \right] \  &\stackrel{t \rightarrow\infty } \longrightarrow &e^{\Gamma_{23} t} \ \left[ \widehat{Y}(0) , \widehat{Z}(0) \right] \\
\left[ \widehat{Z}(t) , \widehat{X}(t) \right] \  &\stackrel{t \rightarrow\infty } \longrightarrow &  e^{- \Gamma_{31} t} \left[ \widehat{Z}(0) , \widehat{X}(0) \right] \nonumber
\eeqa
 For completeness we also present the equations for the closed systems of operators $ \widehat{H}_{2}, \widehat{X} $ 
\beq
\dot{\widehat{H}_{2}} \ + \ b \widehat{H}_{2} \ = \ \sigma \left( 1 \ - \ \frac{b}{2\sigma} \right) \widehat{X}^{2} 
\eeq
and
\beq
\ddot{\widehat{X}} \ + \ \left( 1 + \sigma \right) \dot{\widehat{X}} \ + \ \left( \frac{\widehat{X} \widehat{H}_{2}+
\widehat{H}_{2}\widehat{X}}{2} \ - 
\ \sigma(r-1) \widehat{X} \ + \ \frac{\widehat{X}^{3}}{2} \right) \ = \ 0
\eeq

Currently we are in the process of performing numerical calculations for large symmetric or Hermitian marices 
for the case of weak dissipation where interesting behaviour appears between the quantum and classical overlap regimes. 
For discussions of the Lorenz attractor quantization in the
physical framework of laser instabilities see \cite{es}.

%-------------------------------------------------------------------------%-----------------------------------------------------
\section{ Conclusions-Open Questions \\ Applications}
%-------------------------------------------------------------------------%-----------------------------------------------------
We presented a framework to quantize non-Hamiltonian dissipative flows in $ R^{3}$. 
We applied the method to the 
very interesting and rich example of the Lorenz Strange attractor. 
This we did in order to check the robustness of 
quantum mechanics against the robustness of strange attractors. 
The qualitative behavior for a finite matrix approximations of the quantum system, exhibits interesting results for the case of weak dissipation. This is not the case of a typical Lorenz attractor, 
which exhibits very strong dissipation and undergoes quick 
decoherence. 
We believe the study of quantum dynamical systems with
soft turbulence \cite{es} is useful to probe  
the classical $\leftrightarrow$ quantum regime transition for dissipative dynamical
systems. 

Applications of quantum turbulence  are awaiting for exciting new discoveries 
and new paradigm shifts in the behaviour of quantum matter and radiation in extreme conditions of high temperatures and densities
where the Lagrangian particle concept looses its importance and the Eulerian fluid interpretation 
takes the lead. In all of these, chaotic attractor mechanisms should prevail. Their quantum mechanical survival will be important to our
understanding of quantum turbulence.

In this direction important developments come from recent studies
 of holographic hydrodynamics in its relation with 
black hole horizon physics, where the
problem of quantization of gravity resurfaces from a new hydrodynamic point of view \cite{efo}.

%\end{multicols}
\end{document}